\documentclass[a4paper,reqno,12pt]{amsart}
\usepackage[all]{xy}           
\usepackage{amssymb}           
\usepackage{hyperref}
\usepackage{eucal}
\usepackage{epsfig}

\numberwithin{equation}{section}

\newtheorem{definition}{Definition}[section]
\newtheorem{lemma}[definition]{Lemma}
\newtheorem{theorem}[definition]{Theorem}
\newtheorem{proposition}[definition]{Proposition}

\newtheorem{remarkth}[definition]{Remark}
\newtheorem{example}[definition]{Example}
\newenvironment{remark}{\begin{remarkth}\upshape}{\hfill$\diamond$\end{remarkth}}

\renewcommand{\emph}[1]{{\bfseries\itshape{#1}}}

\makeatletter
\newcommand\prol{\@ifstar{\@proldf}{\@prolpf}}  
\def\@prolpf{\@ifnextchar[{\@prolpf@wrt}{\@prolpf@}}
\def\@prolpf@wrt[#1]#2{\@ifnextchar[{\@prolpf@wrt@at{#1}{#2}}{\@prolpf@wrt@{#1}{#2}}}
\def\@prolpf@wrt@at#1#2[#3]{\prolsymbol^{#1}_{#3}#2}
\def\@prolpf@wrt@#1#2{\prolsymbol^{#1}#2}
\def\@prolpf@#1{\@ifnextchar[{\@prolpf@at{#1}}{\@prolpf@@{#1}}}
\def\@prolpf@at#1[#2]{\prolsymbol_{#2}#1}
\def\@prolpf@@#1{\prolsymbol#1}
\def\@proldf{\@ifnextchar[{\@proldf@wrt}{\@proldf@}}
\def\@proldf@wrt[#1]#2{\@ifnextchar[{\@proldf@wrt@at{#1}{#2}}{\@proldf@wrt@{#1}{#2}}}
\def\@proldf@wrt@at#1#2[#3]{\prolsymbol^{*#1}_{#3}#2}
\def\@proldf@wrt@#1#2{\prolsymbol^{*#1}#2}
\def\@proldf@#1{\@ifnextchar[{\@proldf@at{#1}}{\@proldf@@{#1}}}
\def\@proldf@at#1[#2]{\prolsymbol^*_{#2}#1}
\def\@proldf@@#1{\prolsymbol^*#1}
\def\prolsymbol{\mathcal{T}}
\makeatother

\begin{document}

\title[A Hamilton-Jacobi theory for singular lagrangian systems ]{A Hamilton-Jacobi theory for singular lagrangian systems in the Skinner and Rusk setting}

\author[M. de Le\'on]{Manuel de Le\'on}
\address{Manuel de Le\'on: Instituto de Ciencias Matem\'aticas (CSIC-UAM-UC3M-UCM),
c$\backslash$ Nicol\'as Cabrera,nº 13-15, Campus Cantoblanco,UAM
28049 Madrid, Spain} \email{mdeleon@icmat.es}

\author[D.\ Mart\'{\i}n de Diego]{David Mart\'{\i}n de Diego}
\address{David \ Mart\'{\i}n de Diego:
Instituto de Ciencias Matem\'aticas (CSIC-UAM-UC3M-UCM),
c$\backslash$ Nicol\'as Cabrera,nº 13-15, Campus Cantoblanco,UAM
28049 Madrid, Spain}\email{david.martin@icmat.es}

\author[M. Vaquero]{Miguel Vaquero}
\address{Miguel Vaquero:
Instituto de Ciencias Matem\'aticas (CSIC-UAM-UC3M-UCM),
c$\backslash$ Nicol\'as Cabrera,nº 13-15, Campus Cantoblanco,UAM
28049 Madrid, Spain} \email{miguel.vaquero@icmat.es}


 \subjclass[2000]{}

\begin{abstract}
We develop a Hamilton-Jacobi theory for singular lagrangian systems
in the Skinner-Rusk formalism. Comparisons with the Hamilton-Jacobi problem in the lagrangian and hamiltonian settings are discussed.
\end{abstract}

\maketitle

\tableofcontents

\section{Introduction}

The standard formulation of the Hamilton-Jacobi problem is to find a
function $S(t, q^A)$ (called the {\bf principal function}) such that
\begin{equation}\label{hjj1}
\frac{\partial S}{\partial t} + h(q^A, \frac{\partial S}{\partial
q^A}) = 0,
\end{equation}
where $h=h(q^A, p_A)$ is the hamiltonian function of the system.
If we put $S(t, q^A) = W(q^A) - t E$, where $E$ is a constant, then
$W$ satisfies
\begin{equation}\label{hjj2}
h(q^A, \frac{\partial W}{\partial q^A}) = E;
\end{equation}
$W$ is called the {\bf characteristic function}.

Equations (\ref{hjj1}) and (\ref{hjj2}) are indistinctly referred as
the {\bf Hamilton-Jacobi equation} (see \cite{AM,Arnold,rund}).

This theory works for classical mechanical systems, where the lagrangian function is usually the kinetic energy corresponding to a Riemannian metric on the configuration manifold minus a potential energy. This is the case of the so-called regular lagrangian systems, that have a well-defined hamiltonian counterpart.
The theory has been recently reformulated in a geometrical setting (see \cite{cari,cari2,cari3}) that has permitted its extension to nonholomic mechanical systems \cite{lmm1,lmm3}, and even classical field theories \cite{lmm2,lmm4}.The procedure is based on the comparison of the hamiltonian vector field $X_h$ on the cotangent bundle $T^*Q$ and its projection onto $Q$ via a closed 1-form $\gamma$ on $Q$; the result says that both vector fields are $\gamma$-related if and if the Hamilton-Jacobi equations $d(h\circ \gamma) = 0$ holds.

On the other hand, a Hamilton-Jacobi theory for singular lagrangian systems is far to be accomplished. There were several attempts (\cite{gomis1,gomis2,Rothe}), based on the homogeneization of the given lagrangian, which leads to a new lagrangian system with null energy such that it is possible to discuss the Hamilton-Jacobi equation for the constraints themselves.
The main problem is that, due to the integrability condition for the resultant partial differential equation, one can only consider first class constraints. Therefore, the treatment of the cases when second class constraints appear should be developed by {\it ad hoc} arguments (as in \cite{Rothe}, for instance). Thus, in \cite{gomis1} and \cite{gomis2} the authors only discuss the case of primary constraints.

A more modern discussion on this subject can be found in \cite{cari,sosa}, but these authors only consider the case of primary constraints. More recently, in \cite{hjsingular} it is proposed a Hamilton-Jacobi theory for arbitrary singular systems that works even if the system exhibit secondary constraints. The strategy is to apply the geometric procedure described above in combination with the constraint algoritm developed by M.J. Gotay and J.M. Nester \cite{gotaythesis,gotay2,gotay3,gotay4} and that geometrizes the well-known Dirac theory of constraints \cite{dirac}.

In the present paper we take a different approach, and consider the Skinner and Rusk setting to treat with singular lagrangians \cite{ski,ski2}. Skinner and Rusk have considered a geometrized framework where the velocities and the momenta are independent coordinates. To do this, they considered the dynamics on the Withney sum of $TQ$ (the space of velocities) and $T^*Q$ (the phase space).

Given a lagrangian function $L:TQ\rightarrow \mathbb{R}$ (singular or regular, no matter) one considers the bundle $TQ\oplus T^*Q$ with canonical projections $pr_1:TQ\oplus T^*Q\rightarrow TQ$ and $pr_2:TQ\oplus T^*Q\rightarrow T^*Q$ onto the first and second factors. We then define a function
$D:TQ\oplus T^*Q \longrightarrow  \mathbb{R}$ by $D(X_p,\alpha_p)=\alpha_p(X_p)-L(X_p)$.
In bundle coordinates $(q^A,v^A,p_A)$, $D$ is given by
$D(q^A,v^A,p_A)$ $=v^Ap_A-L(q^A,v^A)$, and it is sometimes refered as the Pontryagin hamiltonian or generalized energy (see \cite{yoshimura}). We can also define a $2$-form $\Omega$ on $TQ\oplus T^*Q$ by $\Omega=pr_2^*(\Omega_Q)$, where $\Omega_Q$ denotes the canonical symplectic 2-form of $T^*Q$.

Then, one discuss the presymplectic system $(TQ\oplus T^*Q, \Omega, dD)$ and obtain the corresponding sequence of constraint submanifolds, which, of course, have a close relation with those obtained by Gotay and Nester on the lagrangian and hamiltonian sides. It should be noticed that this algorithm includes the SODE condition just from the very beginning.

We apply the Hamilton-Jacobi geometric procedure to this presymplectic system and develop the corresponding Hamilton-Jacbi theory. The relation with the Hamilton-Jacobi problems on the lagrangian and hamiltonian sides are extensively discussed.

\section{Notation and background}\label{background}
In this work all manifolds are assumed to be finite dimensional and  $C^{\infty}$. Given a function $f$, the differential at a point $p$
will be indistinctly denoted by $d_pf$ or $df(p)$.

We refer to \cite{Leonrodrigues} for a detailed description of lagrangian and hamiltonian mechanical systems.

Let $Q$ be a differentiable manifold and denote by $TQ$ and $T^*Q$ the tangent and cotangent bundles, and by $\tau_Q:TQ\rightarrow Q$ and $\pi_Q:T^*Q\rightarrow Q $ the respective canonical projections on $Q$.

We introduce two canonical structures on the tangent bundle of a manifold: the vertical endomorfism $S$, and the Liouville vector field $\Delta$. In bundle coordinates, $(q^A,v^A)$, they are respectively given by 
\[
\begin{array}{l}
S = dq^A \otimes \frac{\partial}{\partial v^A},
\\ \noalign{\medskip}
\Delta = v^A \, \frac{\partial}{\partial v^A}.
\end{array}
\]

Let now $L:TQ\rightarrow \mathbb{R}$ be a lagrangian on $TQ$;  we can define the Poincar\'e-Cartan $2$-form and the energy function of $L$ by  
\[
\begin{array}{l}
\Omega_L = - d\theta_L ,  \textrm{ where } \theta_L = S^*(dL),\\ \noalign{\medskip}
E_L = \Delta(L)-L,
\end{array}
\]
which in local coordinates read as
\begin{eqnarray*}
\theta_L &=& \frac{\partial L}{\partial v^A} \, dq^A ,\\
\Omega_L &=& dq^A \wedge d\frac{\partial L}{\partial v^A},\\
E_L &=& v^A \frac{\partial L}{\partial v^A} - L (q, v).
\end{eqnarray*}

We look for vector fields $\xi$ which simultaneously satisfy the equations
\begin{eqnarray}\label{lagrange}
i_{\xi} \, \Omega_L = dE_L \\ \noalign{\medskip}
S \, \xi=\Delta. \label{sode}
\end{eqnarray}
If the lagrangian $L$ is regular, that is, $det(\frac{\partial^2 L}{\partial v^A \partial v^B})\neq 0$, then the form $\Omega_L$ is symplectic ($\Omega_L$ has maximal rank) and there exists a unique vector field $\xi$  on $TQ$ which satisfies the equation \eqref{lagrange}. This vector field automatically satisfies the SODE condition \eqref{sode}.

If the lagrangian is not regular, then $\Omega_L$ is no longer symplectic and equation \eqref{lagrange} has no solution in general and even if there is a solution it is not necesary a SODE. Therefore for a singular lagrangian $L$, $\Omega_L$ is a presymplectic form (that is, the rank is not maximal, althought, for simplicity, it is assumed that it is constant).

We define the Legendre transformation associated to $L$ as the mapping 
\[
\begin{array}{rccl}
FL:& TQ& \longrightarrow &  T^*Q\\ \noalign{\medskip}
& (q^A,v^A)&\rightarrow &FL(q^A,v^A)=(q^A,\frac{\partial L}{\partial v^A}(q^A,v^A)).
\end{array}
\]
From a direct inspection in local coordinates we know that the Legendre transformation is a local diffeomorfism if and only if $L$ is regular.

We can apply the Gotay-Nester-Hinds algorithm of constraints, see
\cite{gotaythesis,gotay2,gotay3}, to the presymplectic system $(TQ, \
\Omega_L, \ dE_L)$ and hence we obtain a sequence of constraint submanifolds
\[
\cdots P_k \hookrightarrow \cdots \hookrightarrow P_2 \hookrightarrow P_ 1= TQ.
\]

Assume that the algorithm stabilizes at some step $k$, say $P_{k+1}=P_k$, which is called the final constraint submanifold, denoted by $P_f=P_k$.

In this paper we will only consider almost regular lagrangians $L:TQ\rightarrow \mathbb{R}$, that is:
\begin{enumerate}
\item $M_1=$Im($\mathbb{F}L$) is a submanifold of $T^*Q$, and
\item $FL: TQ\rightarrow \textrm{Im}(\mathbb{F}L) $ is a surjective submersion of connected fibers.
\end{enumerate}

Under these assumptions, the energy $E_L$ is projected onto a function $h_1:M_1\rightarrow \mathbb{R}$ such that $h_1\circ FL=E_L$ 
\[
\xymatrix{
TQ \ar[rrd]^{FL_1} \ar[rr]^{FL}&& T^*Q \\ 
&& M_1=\textrm{Im}(FL) \ar[u]^{j_1}
}
\]
Here $FL_1$ is the restriction of $FL$ to its image, and $j_1:M_1\rightarrow T^*Q$ is the canonical inclusion.

Next, study the presymplectic system given by $(M_1,\
\Omega_1=j_1^*\Omega_Q,\ dh_1)$, where $\Omega_Q$ is the canonical
symplectic form on $T^*Q$. Therefore, we consider the equation 
\begin{equation}\label{hamilton}
i_{Y}\, \Omega_1=dh_1.
\end{equation}
As above we can apply the presymplectic algorithm and  we obtain a sequence of constraint submanifolds
\[
\cdots M_k \hookrightarrow \cdots \hookrightarrow  M_2 \hookrightarrow M_ 1
\hookrightarrow T^*Q.
\]

It is obvious that
\[ FL(P_i)=M_i \textrm{, for any $i$, }
\]
and, furthermore, the induced mappings
\[
FL_i=FL_{|P_i}:P_i\rightarrow M_i
\]
are surjective submersions, for all $i$.

Hence, both algorithms stabilizes at the same step, say $k$, and then
\[
FL(P_f)=M_f,
\]
and
\[
FL_f:P_f\rightarrow M_f
\]
is a surjective submersion (with the obvious notations).

The following diagram summarizes the above discussion.
\[
\xymatrix{
P_1=TQ\ar[rr]^{FL} \ar[rrd]^{FL_1}  && T^*Q\\
P_2 \ar[u]^{g_2} \ar[rrd]^{FL_2}&& M_1 \ar[u]^{j_1}\\
\vdots & & M_2\ar[u]^{j_2} \\
P_f\ar[u]^{g_f}\ar[rrd]^{FL_f} && \vdots\\
&& M_f\ar[u]^{j_f}
}
\]
where $g_i$ and $j_i$ denote the natural inclusions.

The relation between equations \eqref{lagrange} and \eqref{hamilton} is given by the following theorem.
\begin{proposition}\label{equivalence}
If $\xi\in T_pTQ$ satisfies \eqref{lagrange}, then $TFL(\xi)\in T_{FL(p)}M_1$ satisfies \eqref{hamilton}. Therefore, if $\xi$ is a $FL_f$-projectable solution of \eqref{lagrange}, then its projection $TFL_f(\xi)$ is a solution of \eqref{hamilton}.

Conversely, if $Y$ is a solution of \eqref{lagrange}, then any $FL_f$ projectable vector field on $P_f$ which projects on $Y$, is a solution of \eqref{hamilton}.
\end{proposition}

Next, we shall discuss the SODE problem as it was stated by M.J. Gotay and J.M. Nester \cite{gotaythesis,gotay2}.

The results can be summarized in the following result.

\begin{theorem}\label{sodetheorem}\hfill
\begin{enumerate}

\item If $\xi$ is a $FL_f$-projectable vector field on $P_f$ then for any  $p\in M_f$  there exists a unique point in each fiber $FL_f^{-1}(p)$, denoted by $\eta_{\xi}(p)$ at which $\xi$ is a SODE. The point $\eta_{\xi}(p)$ is given by
\[
\eta_{\xi}(p)=T\tau_Q(\xi(p)).
\]
\item The map 
\[
\begin{array}{rccl}
\beta_{\xi}:& M_f &\longrightarrow & P_f \\ \noalign{\medskip}
& p&\rightarrow & \beta_{\xi}(p)= \eta_{\xi}(p)
\end{array}
\]
is a section of $FL_f:P_f\rightarrow M_f$ and on $\textrm{Im}(\beta_{\xi})$ there exists a unique vector field, denoted by $Y_{\xi}$, which simultaneously satisfies the equations
\[
\begin{array}{lr}
i_{Y_{\xi}} \, \Omega_L=dE_L, &
SY_{\xi}=\Delta.
\end{array}
\] 
\end{enumerate}
\end{theorem}

We will now recall the construction of the  solution of the dynamical equation which simultaneously satisfies the SODE condition. If $Y=(FL_f)_{*}(\xi)$, then $Y$ is a vector field on $M_f$ satisfying $i_{Y} \, \Omega_1=dh_1$. The vector field $Y_{\xi}$ described in \textrm{(ii)} is given by
\[
Y_{\xi}(\beta_{\xi}(p))=T\beta_{\xi}(Y(p)) \textrm{, for all }p\in M_f.
\]

A detailed discussion can be found in \cite{Leonrodrigues,gotaythesis,gotay2,gotay3,gotay6}.

\section{The Skinner and Rusk formalism}
Skinner and Rusk, \cite{ski,ski2}, have considered a geometrized framework where the velocities and the momenta are independent coordinates. Indeed, they considered the dynamics on the Withney sum of $TQ$ (the space of velocities) and $T^*Q$ (the phase space).

In this section we will briefly recall the Skinner and Rusk formalism. 

Let $Q$ be a differentiable manifold and $L:TQ\rightarrow \mathbb{R}$ a lagrangian. We can consider the bundle $TQ\oplus T^*Q$ given by the Withney sum of $\tau_Q:TQ\rightarrow Q$ and $\pi_Q:T^*Q\rightarrow Q $.  We will denote by $pr_1:TQ\oplus T^*Q\rightarrow TQ$ and $pr_2:TQ\oplus T^*Q\rightarrow T^*Q$ the projections onto the first and second factors, and by $pr:TQ\oplus T^*Q\rightarrow Q$ the projection onto $Q$. We then have the following commutative diagram
\begin{equation}\label{diagram}
\xymatrix{
&TQ\oplus T^*Q \ar[dl]^{pr_1}\ar[dd]^{pr} \ar[dr]^{pr_2}& \\
TQ \ar[dr]^{\tau_Q} && T^*Q \ar[dl]^{\pi_Q}\\
&Q
}
\end{equation}

We can define a function
\[
\begin{array}{rccl}
D:& TQ\oplus T^*Q& \longrightarrow & \mathbb{R}\\ \noalign{\medskip}
& (X_p,\alpha_p)&\rightarrow &D(X_p,\alpha_p)=\alpha_p(X_p)-L(X_p).
\end{array}
\]
In bundle coordinates $(q^A,v^A,p_A)$, $D$ is given by
$D(q^A,v^A,p_A)$ $=v^Ap_A-L(q^A,v^A)$. The function $D$ is sometimes refered as the Pontryagin hamiltonian or generalized energy (see \cite{yoshimura}).

We can define a $2$-form $\Omega$ on $TQ\oplus T^*Q$ by $\Omega=pr_2^*(\Omega_Q)$, where $\Omega_Q$ denotes the canonical symplectic 2-form of $T^*Q$.

Next, we can consider the presymplectic system given by $(W_0=TQ\oplus T^*Q,\ \Omega,\ dD)$ and study the equation
\begin{equation}\label{equation}
i_ X\,\Omega=dD,
\end{equation} 
applying the Gotay-Nester-Hinds algorithm of constraints. Hence, we obtain
\[
W_1=\{x\in W_0 \textrm{ such that there exists } X\in T_{x}W_0 \textrm{ satisfying } i_{X}\, \Omega=dD\}.
\]
In canonical coordinates $(q^A,v^A,p_A)$, we have
\[
\begin{array}{l}
\Omega=dq^A\wedge dp_A, \\ \noalign{\medskip}
dD=-\frac{\partial L}{\partial q^A}dq^A+(p_A-\frac{\partial L}{\partial v^A})dv^A+v^Adp_A.\\ \noalign{\medskip}
\end{array}
\]

So, given a tangent vector $X=a^A\frac{\partial }{\partial q^A}+b^A\frac{\partial }{\partial v^A}+c^A\frac{\partial }{\partial p_A}\in T_{(q^A,v^A,p_A)}W_0$ we deduce that
\[
i_{X}\, \Omega=-c^Adq^A+a^Adp_A
\]
and \eqref{equation} is equivalent to the following conditions
\begin{equation}\label{con}
\begin{array}{lr}
a^A=v^A, & \\ \noalign{\medskip}
c^A=-\frac{\partial L}{\partial q^A},&  \\ \noalign{\medskip}
p^A-\frac{\partial L}{\partial v^A}=0, & 1\leq A \leq n.
\end{array}
\end{equation}

Next, we should restrict the dynamics to $W_1=\{(q^A,v^A,p_A)\in W_0 \textrm{ such that } p_A=\frac{\partial L}{\partial v^A}\}$, that is, $W_1=\textrm{graph}(FL)$, where $FL:TQ\rightarrow T^*Q$ has been defined in section \ref{background}.

Accordingly with the Gotay-Nester-Hinds algorithm, a solution $X$ must be tangent to $W_1$. Assume that such $X$ has the local expression
\begin{equation}\label{contan}
\begin{array}{l}
X=\overline{a}^A \frac{\partial}{\partial q^A}+\overline{b}^A \frac{\partial}{\partial v^A}
+ (\frac{\partial^2 L }{\partial v^A\partial q^B }\overline{a}^B+\frac{\partial^2 L }{\partial v^A\partial v^B }\overline{b}^B)\frac{\partial}{\partial p_A}
\end{array}
\end{equation}
Then, taking into account \eqref{con} and \eqref{contan},we deduce
\begin{equation}\label{tan}
\begin{array}{l}
\overline{a}^A=v^A \\ \noalign{\medskip}
\frac{\partial^2 L }{\partial v^A\partial q^B }v^B+\frac{\partial^2 L }{\partial v^A\partial v^B }\overline{b}^B=-\frac{\partial L}{\partial q^A}.
\end{array}
\end{equation}

If there exists such a vector field $X$ tangent to $W_1$, satisfying the above conditions, we have done, and the final constraint manifold $W_f$ is just $W_1$. For instance, if the lagrangian is regular, $det(\frac{\partial^2L}{\partial v^B\partial v^A})\neq 0$, we can compute $\overline{b}^A$ explicitly. If we denote by $C_{A\,B}$ the matrix $C_{A\,B}=\left(\frac{\partial^2L}{\partial v^B\partial v^A}\right)$ and $C^{A\,B}$ its inverse, then
\[
\overline{b}^A=-C^{A\,B}\left(v^A\frac{\partial^2L}{\partial v^B\partial q^A}-\frac{\partial L}{\partial q^A} \right).
\]

Otherwise, we need to continue the process, and then we obtain a sequence of submanifolds
\[
\hdots \hookrightarrow W_k\hookrightarrow\hdots\hookrightarrow W_2\hookrightarrow W_1 \hookrightarrow W_0=TQ\oplus T^*Q. 
\]

If the algorithm stabilizes, that is, there exists $k$ such that $W_k=W_{k+1}$, then $W_k$ is called the final constraint submanifold and denoted by $W_f$.

\section{A Hamilton-Jacobi theory in the Skinner-Rusk setting}

In this section we will develop a Hamilton-Jacobi theory in the
Skinner-Rusk formalism. We will use the same notation introduced in
the previous sections and discuss separately the regular and the singular
cases.

\subsection{The regular case}

Assume that we begin with a regular lagrangian
$L:TQ\rightarrow \mathbb{R}$. Then, 
$W_f=W_1$.

A section of $TQ\oplus T^*Q$ is given by $\sigma= (Z,\gamma)$ where
$Z$ and $\gamma$ are a vector field and a $1$-form on $Q$, respectively. Assume that
$\sigma$ satisfies the following conditions

\begin{enumerate}
\item Im$(\sigma)\subset W_1=\textrm{graph}(FL)$, and
\item $d(pr_2\circ \sigma)=d\gamma=0$.
\end{enumerate}

Then, by the regularity of $L$, we know that there exists a unique
vector field on $W_1$, say $X$, satisfying
\[
i_X \, \Omega=dD,
\]
and then we can define a vector field on $Q$ by
\[
X^{\sigma} (p)=Tpr(X(\sigma(p))), \textrm{ for all } p\in Q.
\]

Now we have the following proposition.
\begin{proposition}\label{rhamiltonjacobi}
Under the previous conditions, $d(D\circ \sigma)=0$ if and only if the
vector fields $X$ and $X^{\sigma}$ are $\sigma$-related.
\end{proposition}

\proof \hfill

``$\Rightarrow$''

Assume that $d(D\circ \sigma)$ $=0$ holds, then we will prove first that $(i_{(X-T\sigma(X^{\sigma}))}\Omega=0)_{|\textrm{Im}(\sigma)}$. 

It is clear that if $x\in \textrm{Im}(\sigma)$ then $T_x(TQ\oplus T^*Q)=T_x\textrm{Im}(\sigma)+V$, where $V$ denotes the vertical bundle of the projection $pr: TQ\oplus T^*Q\rightarrow Q$. We will show that $i_{(X-T\sigma(X^{\sigma}))}\Omega$ anihilates $T_x\textrm{Im}(\sigma)$ and $V$. Indeed, by the definition of $\Omega$, it is obvious that $\Omega$ vanishes acting on two elements of $V$. Since $X-T\sigma(X^{\sigma})$ is vertical, we have
\[
\left(i_{(X-T\sigma(X^{\sigma}))}\Omega\right)(V)=0.
\]

Given $p\in Q$, since $X$ is a solution on $W_1$, we get
\[
(i_{X(p)}\, \Omega)\circ T\sigma(p)=TD({\sigma(p)}) \circ T\sigma(p)=T(D\circ \sigma)(p).
\]

On the other hand, $(i_{T\sigma(X^{\sigma}(p))}\, \Omega) \circ T\sigma(p)=0$ since for any $Y\in T_{p}Q$ we have
\[
\begin{array}{ll}
(i_{T\sigma(X^{\sigma}(p))}\, \Omega) \left( T\sigma(p)(Y)\right)=\Omega(T\sigma(X^{\sigma}(p)),T\sigma(Y))\\ \noalign{\medskip}
=\Omega(T\sigma(X^{\sigma}(p)),T\sigma(Y)) 
=pr_2^*(\Omega_Q)(T\sigma(X^{\sigma}(p)),T\sigma(Y))\\ \noalign{\medskip}=(\Omega_Q)(Tpr_2\circ T\sigma(X^{\sigma}(p)),Tpr_2\circ T\sigma(Y))
=(\Omega_Q)(T\gamma(X^{\sigma}(p)),T\gamma(Y))\\ \noalign{\medskip} =-d\gamma(T\gamma(X^{\sigma}(p)),T\gamma(Y))\\
\noalign{\medskip}=0
\end{array}
\]
and so, we conclude that
\[
\left(i_{(X-T\sigma(X^{\sigma}))}\Omega\right)\left(T\textrm{Im}(\sigma)\right)_{|\textrm{Im}(\sigma)}=0,
\]
which implies 
\[
\begin{array}{l}
\left(i_{(X-T\sigma_f(X^{\sigma}))}\Omega\right)\left(V+T\textrm{Im}(\sigma)\right)_{|\textrm{Im}(\sigma)}\\ \noalign{\medskip}=\left(i_{(X-T\sigma_f(X^{\sigma}))}\Omega\right)\left(T(TQ\oplus T^*Q)\right)_{|\textrm{Im}(\sigma)}=0.
\end{array}
\]
Therefore $(X-T\sigma(X^{\sigma}))\in \ker(\Omega)$. This means that $i_{(X-T\sigma(X^{\sigma}))} \, \Omega=0$, and hence $\iota^*\left(i_{(X-T\sigma(X^{\sigma}))} \, \Omega\right)$ $=i_{(X-T\sigma(X^{\sigma}))} \,(i^* \Omega)=0$, where $\iota:W_1\rightarrow W_0$ is the inclusion.

It is not hard to see, that if $L$ is regular then $i^*\Omega$ is symplectic and so $\left(X=T\sigma(X^{\sigma})\right)_{|\textrm{Im}(\sigma)}$.

``$\Leftarrow$''
Since $\left((i_{(X-T\sigma(X^{\sigma}))}\Omega)\circ T\sigma=d(D\circ \sigma)\right)$, if $X=T\sigma(X^{\sigma})$, then $d(D\circ \sigma)=0$.
\qed

\subsection{The singular case}

Assume now that 
$L:TQ\rightarrow \mathbb{R}$ is an almost regular singular lagrangian.

Suppose that the algorithm of Gotay-Nester-Hinds applied to $(W_0=TQ\oplus T^*Q,\ \Omega,\ dD)$  stabilizes at a final constraint submanifold $W_f$. By construction, there exists at least one vector field $X$ on $W_f$ such that
\[
(i_X\, \Omega=dD)_{|W_f}
\]

We need some regularity conditions, thus we will also assume that $Q_i=pr(W_i)$ are submanifolds and that $pr_i=pr_{|W_i}:W_i\rightarrow Q_i$ are submersions.

A section of $pr: TQ\oplus T^*Q\rightarrow Q$ is given by $\sigma=(Z,\gamma)$, where $Z$ and $\gamma$ are respectively a vector field and a $1$-form on $Q$. We will denote by $\sigma_f$ the restriction of $\sigma$ to $Q_f=pr(W_f)$ of $\sigma$. Suppose that $\sigma$ verifies the following conditions:
\begin{enumerate}
\item Im$(\sigma)\subset W_1$.
\item Im$(\sigma_f)\subset W_f$.
\item $d(pr_2\circ \sigma)=d\gamma=0$, that is, $\gamma$ is closed.
\end{enumerate}

Using $\sigma$ we can define a vector field on $Q_f$ by
\[
X^{\sigma}(p)=Tpr(X(\sigma_f(p))), \quad p\in Q_f.
\]

The construction is illustrated in the following diagram
\[
\xymatrix{
W_0 \ar@/^/[dd]^{pr} & &\ar[ll] W_f\ar@/^/[dd]^{pr_f} \ar[rr]^{X}&& TW_f\ar[dd]^{Tpr_f} \\ \noalign{\medskip}
\\ 
Q \ar@/^/[uu]^{\sigma} && \ar[ll] Q_f \ar@/^/[uu]^{\sigma_f}  \ar[rr]^{X^{\sigma}} &&  TQ_f.
}
\]
The relation between $T\sigma_f(X^{\sigma})$ and $X$ is shown in the following theorem.
\begin{proposition}\label{hamiltonjacobi}
 The conditions 
$$
d(D\circ \sigma)_{|Q_f}=0
$$ 
and 
$$
\left(X-T\sigma_f(X^{\sigma})\in \ker(\Omega)\right)_{|\textrm{Im}(\sigma_f)}
$$
are equivalent.
\end{proposition}
\proof 
The proof follows by similar arguments as in 
 Proposition \ref{rhamiltonjacobi}.
\qed

\begin{definition} A section $\sigma$ of $TQ\oplus T^*Q$, $\sigma=(Z,\gamma)$, satisfying the following conditions
\begin{enumerate}
\item Im$(\sigma)\subset W_1$.
\item Im$(\sigma_f)\subset W_f$.
\item $d(pr_2\circ \sigma)=d\gamma=0$.
\item  $d(D\circ \sigma)_{|Q_f}$ $=0$
\end{enumerate}
 will be called a solution of \textbf{the  Hamilton-Jacobi problem for the lagrangian $L$ in the Skinner-Rusk setting}.
\end{definition}

\begin{remark}
The last proposition says that $T\sigma_f(X^{\sigma})$ is a vector field along $\textrm{Im}(\sigma_f)$ which is also a solution of the equation \eqref{equation}. So if we find an integral curve $c(t)$ of $X^{\sigma}$ on $Q_f$, then $(\sigma_f\circ c)(t)$ is an integral curve of a solution of \eqref{equation}.
\end{remark}

\begin{remark}
The natural question is if $X$ and $X^{\sigma}$ are $\sigma_f$-related in the singular case, as it happens in the standard Hamilton-Jacobi theory, see \cite{hjsingular}. The answer is that, as we discussed later (section \ref{conclusion}), in some cases the fields are not necessarily $\sigma_f$-related. 
\end{remark}

\section{Comparison with the Hamiltonian and lagrangian settings}
In the previous section we have developed a Hamilton-Jacobi theory in the Skinner-Rusk setting. The Skinner-Rusk formalism  unifies lagrangian and hamiltonian formalisms, so we would like to relate the present Hamilton-Jacobi theory to the corresponding ones for the two formalisms (see \cite{hjsingular}).

\subsection{The hamiltonian setting}
\subsubsection{The \textbf{regular case}}
If the lagrangian, $L$, is regular, that is, $FL$ is a local
diffeomorfism, then we can define locally a hamiltonian function $h:T^*Q \rightarrow \mathbb{R}$ by $h=E_L\circ FL^{-1}$. Let
us now assume that the lagrangian is hyperregular, that is,
$FL$ is a global diffeomorfism and $h$ is globally defined. Denote by $X_h$
the corresponding hamiltonian vector field
\[
i_{X_h}\, \Omega_Q=dh.
\]
Let $\gamma$ be a closed $1$-form on $Q$; then we can define a vector field on $Q$ by
\[
X^{\gamma}(p)=T\pi_Q(X_h(\gamma(p)))\textrm{ for all }p\in Q.
\]

Then we have the following Hamilton-Jacobi theorem.

\begin{proposition}\label{hamiltonjacobiregular}
The vector fields $X$ and $X^{\gamma}$ are $\gamma$-related if and
only if $d(h\circ \gamma)=0$.
\end{proposition}
\proof 
For a proof see \cite{AM}.
\qed

\subsubsection{The singular case}
Since we are considering  an almost regular lagrangian $L:TQ\rightarrow \mathbb{R}$, then we can apply the Dirac theory of constraints developed in Section \ref{background}. 

We have to study the presymplectic system given by $(M_1,\ \Omega_1=j_1^*\Omega_Q,\ dh_1)$, where $j_1:M_1\rightarrow T^*Q$ is the inclusion and $h_1$ is defined implicitly by $h_1\circ FL= E_L$.

If we apply the Gotay-Nester-Hinds algorithm, we obtain a sequence
\[
\cdots M_k \hookrightarrow \cdots \hookrightarrow M_2 \hookrightarrow M_ 1
\hookrightarrow T^*Q;
\]
assume that we obtain a final constraint submanifold, denoted by $M_{f}$. We also assume that $Q_i=\pi_Q(M_i)$ are submanifolds and that $\pi_i={\pi_Q}_{|M_i}:M_i\rightarrow Q_i$ are submersions.

\begin{remark} It is important to notice that the algorithm of Gotay-Nester-Hinds  applied to the same lagrangian in the Skinner-Rusk setting and in the corresponding hamiltonian setting does not necessary stop at the same level. For example, the lagrangian given by $L(q^1,q^2,v^1,v^2)=v^1 \, q^2$ produces the two presymplectic systems $(M_1,\Omega_1,\linebreak dh_1)$ and $(W_0=TQ\oplus T^*Q, \Omega, dD)$. The first algorithm stabilizes in $k=1$, but the second one does in $k=2$.

\end{remark}
Let  $\gamma$ be a $1$-form on $Q$ satisfying the following conditions:

\begin{enumerate}
\item Im$(\gamma)\subset M_1$.
\item Im$(\gamma_{f})\subset M_{f}$, where $\gamma_{f}$ denotes the restriction to $Q_{f}$ of $\gamma$.
\item $d\gamma=0$.
\end{enumerate}

Then, if $Y$ is a vector field on $M_{f}$ solving the equation $i_Y \, \Omega_1=dh_1$, we can construct the vector field $Y^{\gamma}$ on $Q_{f}$ biven by
\[
Y^{\gamma}(p)=T\pi_Q(Y(\gamma_{f}(p))), \quad  \textrm{ for each }p\in Q_{f} 
\]
and obtain an analogous of theorem \ref{hamiltonjacobi} (notice that in this case we can ensure that the vector fields are $\gamma_{f}$-related, see \cite{hjsingular} for the details).

\begin{proposition}\label{hhamiltonjacobi}
We have
\[
d(h_1\circ \gamma)_{|Q_{f}}=0 \Leftrightarrow Y \textrm{ and } Y^{\gamma} \textrm{ are $\gamma_{f}$-related}.
\]
\end{proposition}
\proof
Given $q\in Q_{f}$, we have
\[
\begin{array}{l}
\left(i_{\left(Y(\gamma(q))-T_{q}\gamma_{f}(Y^{\gamma}(q))\right)}\, \Omega_1\right) \circ T_{q}\gamma= i_{Y(\gamma(q))}\, \Omega_1\circ T_{q}\gamma-i_{T_{q}\gamma_{f}(Y^{\gamma}(q))}\, \Omega_1\circ T_{q}\gamma\\ \noalign{\medskip}
=d_{\gamma_{f}(q)}h_1\circ T_{q}\gamma=d_{q}(h_1\circ \gamma)
\end{array}
\]

\noindent where we have $T_{q}\gamma_{f}(Y^{\gamma})=T_{q}\gamma(Y^{\gamma})$ and  
\[
\begin{array}{l}
i_{T_{q}\gamma_{f}(Y^{\gamma}(q))}\, \Omega_1\circ T_{q}\gamma(Y(q))= \Omega_1(T_{q}\gamma(Y^{\gamma}), T_{q}\gamma(Y(q)))\\ \noalign{\medskip}
=(\gamma^*\Omega_1)(Y^{\gamma}(q),Y(q))=d\gamma(Y^{\gamma}(q),Y(q))=0,
\end{array}
\]
for all $Y_{q}\in T_{q}Q$.

The previous discussion can be applied to every point $q\in Q_{f}$; therefore, taking into account that $\Omega_1$  vanishes acting on two vertical tangent vectors, we can deduce the following
\[
Y-T\gamma_{f}(Y^\gamma) \in \ker( \Omega_1) \Leftrightarrow d(h_1 \circ
\gamma)_{|Q_{f}} = 0.
\]

As we did before, we will see that  $Y$ and $Y^{\gamma}$ are $\gamma_{f}$ related.

Remember  that for any point $p$ of $M_1$ we have a decomposition
\[
T_p(T^*Q)=T_pM_1+V_p(T^*Q),
\]
where $V(T^*Q)$ denotes as above the space of vertical tangent  vectors on $p$.

Since $Y-T\gamma_{f}(Y^{\gamma})$ is vertical at the points of $\textrm{Im}(\gamma_{f})$, given any $U\in V_p$, $p\in \textrm{Im}(\gamma_{f})$, then 
\[
\Omega_Q(Y-T\gamma(Y^{\gamma}),U)=0
\]

Now, given $U\in T_pM_1$ we get
\[
\Omega_Q(Y-T\gamma_{f}(Y^{\gamma}),U)=\Omega_1(Y-T\gamma_{f}(Y^{\gamma}),U)=0
\]
because $(Y-T\gamma_{f}(Y^{\gamma}))\in \ker ( \Omega_1)$, and hence $\Omega_Q(Y-T\gamma_{f}(Y^{\gamma}),Z)$ $=0$ for any tangent vector $Z\in T_p(T^*Q)$ at any point of $\textrm{Im}( \gamma_{f})$. Since $\Omega_Q$ is non-degenerate, we deduce that $Y=T\gamma_{f}(Y^{\gamma})$ along $\textrm{Im}( \gamma_{f})$.

\qed

\begin{definition} A $1$-form $\gamma$ satisfying the previous conditions will be called a solution of the \textbf{Hamilton-Jacobi problem for $L$ in the hamiltonian setting}.
\end{definition}

We are now going to relate the Hamilton-Jacobi problem in the Skinner-Rusk setting and the corresponding one in the hamiltonian setting. First, the following result gives the relation between $W_i$ and $M_i$, and also a relation between solutions of equations \eqref{hamilton} and \eqref{equation}.
\begin{lemma}\label{ham}
\begin{enumerate} \hfill
\item If $X\in T_pW_1$ satisfies $i_X \, \Omega=dD$, then $X_2=Tpr_2(X)$ $\in T_{pr_2(p)}M_1$ satisfies $i_{X_2} \, \Omega =dh_1$.
\item For each step $k$ of the constraint algorithms applied to the presymplectic systems $(M_1,\ \Omega_1,\ dh_1)$ and $(W_0=TQ\oplus T^*Q,\ \\ \Omega,\ dD)$ we have
\[
pr_2(W_k)\subset M_k,
\]
and, if we denote the respective final constraint submanifolds by $W_f$ and $M_{f}$, then
\[
pr_2(W_f)=M_{f}.
\]
\item We have $pr(W_f)=\pi_Q(M_{f})=Q_{f}$.
\end{enumerate}
\end{lemma}
\proof \hfill

(i) Recall that a vector $\xi\in T_{(q^A,v^A)}TQ$, $\xi=u^A\frac{\partial}{\partial q^A}+w^A\frac{\partial}{\partial v^A}$ satisfies $i_{\xi}\, \Omega_L=dE_L$ iff 
\[
\begin{array}{l}
\frac{\partial^2 L}{\partial v^A \partial v^B}(v^B-u^B)=0 \\  \noalign{\medskip}
\frac{\partial^2 L}{\partial v^A \partial v^B}u^B+\frac{\partial^2 L}{\partial v^A \partial q^B}w^B-\frac{\partial L}{\partial q^A}=\frac{\partial^2 L}{\partial v^B \partial q^A}(v^B-u^B)
\end{array}
\]

If $X\in T_pW_1$ verifies $i_X \, \Omega=dD$, then  $X$ has the expression \eqref{contan} and satisfies \eqref{tan}. So,  it is clear that $X_1=Tpr_1(X)$ satisfies $i_{X_1}\, \Omega_L=dE_L$. Since $X$ is tangent to $W_1$, $X_2=Tpr_2(X)=TFL\circ Tpr_1(X)=TFL(X_1)$ and using Proposition \ref{equivalence} we can conclude that  $i_{X_2} \, \Omega_1=dh_1$.

(ii) It will be proved by induction.

For $k=1$ we have $pr_2(W_1)=M_1$ since $W_1=\textrm{graph}(FL)$.

Assume that $pr_2(W_k)\subset M_k$. Then
\[
\begin{array}{l}
W_{k+1}=\{x\in W_{k} \textrm{ such that there exists } X\in T_{x}W_{k}  \textrm{ satisfying } i_{X}\, \Omega=dD\} \\ \noalign{\medskip}
M_{k+1}=\{y\in M_{k} \textrm{ such that there exists } Y\in T_{y}M_{k} \textrm{ satisfying } i_{Y}\, \Omega_1=dh_1\}.
\end{array}
\]

If $x\in W_{k+1}$, then there exists $ X\in T_{x}W_{k}$, satisfying $i_{X}\, \Omega=dD$. Since $pr_2(W_k)\subset M_k$, $Tpr_2(X)\in TM_k$ and by (i) $i_{Tpr_2(X)}\, \Omega_1=dh_1$. Thus, we have proved that $pr_2(x)\in M_{k+1}$ and that $pr_2(W_k)\subset M_k$.

To prove that $pr_2(W_f)=M_{f}$, take a solution $Y$ of equation \ref{hamilton} on $M_f$. Then we can construct a vector field $\xi$ on $P_{f}$ which is $FL_f$-related with $Y$, and using Theorem \ref{sodetheorem} we obtain a vector field $Y_{\xi}$ along the image of the section $\beta_{\xi}$ which satisfies \eqref{lagrange} and \eqref{sode}. We can construct the map
\[
\begin{array}{rccl}
\overline{\beta}_{\xi}:& M_f& \longrightarrow & TQ\oplus T^*Q \\ 
&(q^A,p^A)& \rightarrow & (\beta_{\xi}(q^A,p_ A),(q^A,p_A)).
\end{array}
\]

It is easy to see, that the vector field $T\overline{\beta}_{\xi}(Y)$ on Im($\overline{\beta}_{\xi}$) is a solution of \eqref{equation}. By the maximality of the final constraint manifold $W_f$, we can conclude that Im($\overline{\beta}_{\xi}$)$\subset W_f$, but $M_{f}=pr_2(\textrm{Im}(\overline{\beta}_{\xi}))\subset pr_2(W_f)\subset M_{f}$ and then the result follows.

(iii) It is a direct consequence of (ii) and the commutativity of diagram \eqref{diagram}.
\qed

\vspace{0.3cm}

A solution of the Hamilton-Jacobi problem as stated in the previous section is given by a section $\sigma$ of  $TQ\oplus T^*Q$, so $\sigma=(Z,\gamma)$, where $Z$ and $\gamma$ are a vector field and a $1$-form on $Q$, respectively.

We will see that $\gamma$ satisfies the  Hamilton-Jacobi problem in the hamiltonian sense.

From the fact that $\sigma$ is a solution of the Hamilton-Jacobi problem in the Skinner-Rusk setting, we deduce:
\begin{enumerate}
\item Since Im$(\sigma)\subset W_1$, then Im$(\gamma)=pr_2(Im(\sigma))\subset pr_2(W_1)$  $=M_1$.
\item Since Im$(\sigma_f)\subset W_f$, then Im$(\gamma_f)=pr_2 (Im(\sigma_f))\subset pr_2(W_f)$ $=M_f$.
\item Since $d(pr_2\circ \sigma)=d\gamma=0$, then $\gamma$ is closed.
\item Since Im$(\sigma)\subset W_1$, then $D\circ \sigma=h_1\circ \gamma$ and then, using that  $d(D\circ \sigma)_{|Q_f}=0$, we finally get $d(h_1\circ \gamma)_{|Q_f}=0$.
\end{enumerate}

On the other hand, given a vector field $X$ on $W_f$ which is a  solution of \eqref{equation}, we can obtain a solution of \eqref{hamilton} along Im$(\gamma_f)$ by defining
\[
X_2(\gamma_f(p))=Tpr_2(X(\sigma_f(p)))  \textrm{, for all }p\in Q_f.
\]
Now, from Lemma \ref{ham} it follows that $X_2$ is a solution of \eqref{hamilton}.

As above we can construct the projected vector field on $Q_f$,  by putting
\[
X_2^{\gamma}(p)=T\pi_f(X_2(\gamma_f(p))) \textrm{, for all }p\in Q_f.
\]

\begin{remark}
By the commutativity of the diagram \eqref{diagram} we deduce that $pr=\pi_Q\circ pr_2$, and in consequence we have
\[
X^{\sigma}(p)=Tpr(X(\sigma_f(p)))=T\pi_Q\circ pr_2(X(\sigma_f(p)))=T\pi_f(X_2(\gamma_f(p)))
\]
$\textrm{ for all }p\in Q_f$, and so, $X^{\sigma}=X_2^{\gamma}$.
\end{remark}

Summarizing the above discussion, we can conclude that it is possible to relate the Hamilton-Jacobi theory in the Skinner-Rusk setting to the Hamilton-Jacobi theory on $T^*Q$. In this case the vector fields $X_2$ and $X_2^{\gamma}$ are $\gamma_f$-related. 

\subsection{ The lagrangian setting}
In this section we will relate the Hamil\-ton-Jacobi theory developed in the Skinner-Rusk setting with the corresponding one on the lagrangian side

\subsubsection{The regular case}
If the lagrangian $L$ is regular, then we have a
symplectic system given by $(TQ, \ \Omega_L, \ E_L)$. Then there
exists a unique solution $\xi$ of the equation \ref{lagrange} which
automatically satisfies the SODE condition. 

Given $Z$ a vector field on $Q$ such that $Z^*\Omega_L=0$ we can
define the following vector field on $Q$
\[
\xi^{Z}(p)=T\tau_Q(\xi(Z(p)))\textrm{ for all }p\in Q
\]
and obtain the following result.
\begin{proposition}
Under the previous conditions, the vector fields $\xi$ and $\xi^{Z}$
are $Z$-related if and only if $d(E_L\circ Z)=0$.
\end{proposition}
\proof
The proof is a consequence of Proposition \ref{hamiltonjacobiregular}.
\qed

\subsubsection{The singular case}
In this case, we will discuss the presymplectic system given by $(TQ,\Omega_L,dE_L)$. 
Applying the Gotay-Nester-Hinds algorithm we obtain a sequence of submanifolds
\[
\cdots P_k \hookrightarrow \cdots \hookrightarrow P_2 \hookrightarrow P_ 1= TQ.
\] 
We also assume that $Q_i=\tau_Q(P_i)$ are submanifolds and that $\tau_i={pr_Q}_{|P_i}:P_i\rightarrow Q_i$ are submersions, for any index $i$.

Remember that the algorithm of Gotay-Nester-Hinds applied to the presymplectic systems $(M_1,\ \Omega_1,\ dh_1)$ and $(TQ,\ \Omega_L,\ dE_L)$ stop at the same step, so we will denote the final constraint manifold of the system $(TQ,\Omega_L,dE_L)$ by $P_{f}$.

Let $Z$ be a vector field on $Q$  satisfying the following properties:

\begin{enumerate}
\item Im$(Z_{f})\subset P_{f}$, where $Z_{f}$ denotes the restriction of $Z$ to $Q_{f}$.
\item $Z^*\Omega_L=0$.
\end{enumerate}

Then, if $\xi$ is a vector field on $P_{f}$ solving the equation $i_{\xi} \, \Omega_L=dE_L$, we can construct the vector field $\xi^{Z}$ on $Q_{f}$ by
\[
\xi^{Z}(p)=T\tau_Q(\xi(Z_{f}(p))),  \textrm{ for all }p\in Q_f.
\]
Now, we can develop the corresponding Hamilton-Jacobi theory in the lagrangian setting.

\begin{proposition}
Under the above hypothesis for $Z$ we have
\[
d(E_L\circ Z)_{|Q_{f}}=0 \Leftrightarrow \left( \xi-TZ_{f}(\xi^{Z})\right)\in \ker(\Omega_L).
\]
\end{proposition}

\proof 
``$\Rightarrow$''

Assume that $d(E_L\circ Z)_{|Q_{f}}=0$ holds, then we will prove that 
$$
(i_{\left( \xi-TZ_{f}(\xi^{Z})\right)}\Omega_L=0)_{|\textrm{Im}(Z_{f})}.
$$ 

For any $x\in \textrm{Im}(Z_{f})$ we have the decomposition $T_x(TQ)=T_x\textrm{Im}(Z)+V(TQ)$, where $V(TQ)$ denotes the vertical bundle of the projection $\tau_Q :TQ\rightarrow Q$.

Since $\Omega_L$ vanishes acting on two elements of $V(TQ)$ and $\xi-TZ_{f}(\xi^{Z})$ is vertical, we have
\[
\left(i_{\left( \xi-TZ_{f}(\xi^{Z})\right)}\Omega\right)(V(TQ))=0
\]

Since $\xi$ is a solution along Im$(Z_{f})$,  we have 
\[
\left(i_{\xi(p)}\, \Omega_L\right)\circ T\sigma(p)=d_{Z(p)}E_L\circ TZ_{f}(p)=d_{p}(E_L\circ Z)
\]
for any $p\in Q_{f}$.

On the other hand, $(i_{TZ_{f}(\xi^{Z}(p))}\, \Omega_L) \circ TZ(p)=0$, since for any $Y\in T_{p}Q$ we get
\[
\begin{array}{ll}
\left(i_{\left( \xi-TZ_{f}(\xi^{Z})\right)}\, \Omega_L\right) \circ TZ(p)(Y)=\Omega_L(TZ_{f}(\xi^{Z}(p)),TZ(Y))
\\ \noalign{\medskip} =\left(Z^*\Omega_L\right)((\xi^{Z},Y) 
=-d\gamma(\xi^{Z},Y)=0
\\ \noalign{\medskip}

\end{array}
\]
and so we can conclude that
\[
\left(i_{\left( \xi-TZ_{f}(\xi^{Z})\right)}\Omega_L\right)(T\textrm{Im}(Z))=0
\]

``$\Leftarrow$''

Since $i_{\left( \xi-TZ_{f}(\xi^{Z})\right)}\Omega_L=d(E_L\circ Z)$, if $( \xi-TZ_{f}(\xi^{Z}))\in \ker(\Omega_L)$, then $d(E_L\circ Z)_{|Q_{f}}=0$
\qed

\begin{definition} A vector field on $Q$, $Z$ satsifying the previous conditions will be called a solution of the \textbf{ Hamilton-Jacobi problem for $L$ in the lagrangian setting}.
\end{definition}

The vector fields $\xi$ and $\xi^Z$ are not necessarily related as the next example shows.

\begin{example}\label{example}{\rm
Let $L:T\mathbb{R}^2\rightarrow \mathbb{R}$ be the lagrangian given by
 \[L(q^1,q^2,v^1,v^2)=q^1\,v^2+q^2\,v^1\]

We have
\[
\begin{array}{l}
FL(q^1,q^2,v^1,v^2)=(q^1,q^2,q^2,q^1), \\ \noalign{\medskip}
E_L(q^1,q^2,v^1,v^2)=q^1\,v^2+q^2\,v^1-q^1\,v^2-q^2\,v^1=0, \\ \noalign{\medskip}
\Omega_L=0,
\end{array}
\]
so every vector field $\xi$ on $T\mathbb{R}^2$ satifies
\[
i_{\xi}\, \Omega_L=dE_L.
\]
Therefore, the algorithm of Gotay-Nester-Hinds stabilizes at the first step, and $P_{f}=P_1=TQ$.

Moreover, every vector field $Z$ on $\mathbb{R} ^2$ is a solution of the  Hamilton-Jacobi problem, since $E_L\circ Z=0$ and $Z^*\Omega_L=0$.

Let $\xi$ be the solution satisfying the SODE condition given by
\[
\xi(q^1,q^2,v^1,v^2)=v^1\frac{\partial}{\partial q^1}+v^2\frac{\partial}{\partial q^2}+\frac{\partial}{\partial v^1}+\frac{\partial}{\partial v^2}
\]

Let $Z$ be
\[
Z(q^1,q^2)=\frac{\partial}{\partial q^1}+\frac{\partial}{\partial q^2}
\]

An easy computation shows that
\[
TZ(Z(q^1,q^2))=\frac{\partial}{\partial q^1}+\frac{\partial}{\partial q^2},
\]
but
\[
\xi(Z(q^1,q^2))=\frac{\partial}{\partial q^1}+\frac{\partial}{\partial q^2}+\frac{\partial}{\partial v^1}+\frac{\partial}{\partial v^2}\neq TZ(Z(q^1,q^2)).
\]
Thus, the vector fields $\xi$ and $\xi^Z$ are not $Z$-related.
}
\end{example}

Next we will show that a solution of the Hamilton-Jacobi problem in the Skinner-Rusk formalism induces a solution of the Hamil\-ton-Jacobi theory in the lagrangian setting.

The following lemma is analogous to Lemma \ref{ham}.

\begin{lemma}\label{lag}
\begin{enumerate}\hfill
\item If $X\in T_pW_1$ satisfies $i_X \, \Omega=dD$, then $X_2=Tpr_1(X)$ satisfies $i_{X_1} \, \Omega_L =dE_L$ and the SODE condition \eqref{sode}.
\item For each step $k$ of the constraint algorithm applied to the presymplectic systems $(M_1,\ \Omega_1,\ dh_1)$ and $(W_0=TQ\oplus T^*Q,\  \Omega,\ dD)$, we have
\[
pr_2(W_k)\subset P_k
\]
\item We have $pr(W_f)=\tau_Q(P_{f})$
\end{enumerate}
\end{lemma}
\proof \hfill

(i) and (ii) are proved using similar arguments to that in Lemma \ref{ham}.

(iii) Since the following diagram 
\[
\xymatrix{
P_{f} \ar[ddr]^{\tau_f}\ar[rrd]^{FL_{f}} && \\
                  && M_{f} \ar[dl]^{\pi_{f}}\\
&Q_f&
}
\]
is commutative, and $FL_{f}$ is a surjective submersion, we deduce that $\pi_Q(M_{f})=\tau_Q(P_{f})$. By Lemma \ref{ham} (iii), we obtain $\pi_Q(M_{f})=pr(W_f)$, and the result follows. The situation can be summarized in the following commutative diagram
\[
\xymatrix{
& W_f \ar@/_/[ddl]_{pr_1} \ar@/^/[ddr]^{pr_2}\ar[d]^{pr}& \\
& {\tiny pr(W_f)=\tau_Q(P_{f})} &  \\
P_{f} \ar[ur]_{\tau_Q} \ar[rr]^{FL}&&\ar[ul]^{\pi_Q} M_{f}
}
\]
\qed

If  $\sigma=(Z,\gamma)$ is a solution of the Hamilton-Jacobi problem, we deduce the following results:

\begin{enumerate}
\item Since Im$(\sigma_f)\subset W_f$, then $pr_1($Im$(\sigma_f))\subset pr_1(W_f)\subset P_{f}$.
\item  We have $Z^*\Omega_L=0$, since $Z^*\Omega_L=Z^*(d\theta_L)=d(Z^*\theta_L)=d(FL(Z))=d\gamma=0$.
\item Since Im$(\sigma)\subset W_1$, then $D\circ \sigma(p)=E_L\circ Z(p)$ and, because $d(D\circ \sigma)_{|Q_f}=0$, then $d(E_L\circ Z)_{|Q_f}=0$.
\end{enumerate}

Now, given a solution $X$ of \eqref{equation}, we can obtain a solution of \eqref{lagrange} along Im$(Z_f)$ using Lemma \ref{lag}, and putting
\[
X_1(Z_f(p))=Tpr_1(X(\sigma(p))),\textrm{ for all }p\in Q_f
\]
We can also define the vector field on $Q_f$ given by
\[
X_1^{Z}(p)=T\tau_Q(X_1(Z_f(p))).
\]

The vector fields $X_1$ and $X_1^Z$ are not  $Z_f$-related in general, as we have proved in example \ref{example}.

\begin{remark}
By the commutativity of diagram \eqref{diagram} we have $pr$ $=\tau_Q\circ pr_1$ and hence

$$
X^{\sigma}(p)=Tpr(X(\sigma_f(p)))=T\tau_Q\circ pr_1(X(Z_f(p)))
=T\tau_Q(X_1(Z_f(p))),
$$
for all $p\in Q_f$, and so $X^{\sigma}=X_1^{Z}$.

Moreover, since $X_1$ satisfies the SODE condition, then 
$$
X_1^Z(p)=T\tau_Q(X_1(Z(p)))=\tau_{TQ}(X_1(Z(p)))=Z(p)=Z_f(p),
$$
and we have
\[
X^{\sigma}=X_1^{Z}=X_2^{\gamma}=Z_f.
\]

Note that this means that we only need to compute $X_2^{\gamma}$ to obtain $Z_f$.
\end{remark}

\section{Final considerations}\label{conclusion}

In the last section we show that a solution of the Hamilton-Jacobi problem in the Skinner-Rusk setting, $\sigma=(Z,\gamma)$, gives a solution of the  Hamilton-Jacobi problem in the lagrangian and hamiltonian settings ($Z$ and  $\gamma$ respectively). A solution of the equation \ref{equation} along Im$(\sigma)$ can be also projected to  solutions of \ref{lagrange} and \ref{hamilton} along Im$(Z)$ and Im$(\gamma)$, denoted respectively by $X_1$ and $X_2$. 

If we take a vector field $X$ solution of the equation \eqref{equation} on $W_f$, using $\sigma$ we can compute $X^{\sigma}$. Now we can easily conclude that the vector fields $X$ and $X^{\gamma}$ are  $\sigma_f$ related iff the corresponding vector fields $X_1$ and $X_1^Z$ are $Z_f$ related in the lagrangian setting.

To illustrate the above results we revisite example \ref{example} in the Skinner-Rusk setting and apply the corresponding Hamilton-Jacobi theory.

\begin{example}{\rm
Consider the lagrangian given in Example \ref{example}
 \[
L(q^1,q^2,v^1,v^2)=q^1\,v^2+q^2\,v^1.
\]

Then, on $T\mathbb{R}^2\oplus T^*\mathbb{R}^2$ we have 
\[
D(q^1,q^2,v^1,v^2,p_1,p_2)=v^1p_1 +v^2p_2+v^1q^2+v^2q^1,
\]
and hence
\begin{equation}\label{differential}
\begin{array}{ll}
dD(q^1,q^2,v^1,v^2,p_1,p_2)=&-v^2dq^1-v^1dq^2+(p_1-q^2)dv^1 \\\noalign{\medskip} & +(p_1-q^1)dv^2+v^1dp_1+v^2dp_2
\end{array}
\end{equation}

Recall that we must compute 
\[
\begin{array}{ll}
W_1=&\{(q^A,v^A,p_A) \textrm{ such that there exists } X\in T_{(q^A,v^A,p_A)}T\mathbb{R}^2\oplus T^*\mathbb{R}^2 \\ \noalign{\medskip} & \textrm{ satisfying } i_{X}\, \Omega=dD\}.
\end{array}
\]

If
\begin{equation}\label{X}
X=a^1 \frac{\partial}{\partial q^1}+a^2 \frac{\partial}{\partial q^2}+b^1 \frac{\partial}{\partial v^1}+b^2 \frac{\partial}{\partial v^2}+c^1 \frac{\partial}{\partial p_1}+c^2 \frac{\partial}{\partial p_2}
\end{equation}
then
\begin{equation}\label{contraction}
i_{X}\, \Omega=-c^1dq^1-c^2dq^2+a^1dp_1+a^2dp_2
\end{equation}
and so
\begin{equation}\label{conditions}
\begin{array}{llllll}
a^1=v^1, &
a^2=v^2, &
c^1=v^2, &
c^2=v^1,&
p_1-q^2=0, &
p_2-q^1=0
\end{array}
\end{equation}
must hold.

Therefore, $W_1=\{(q^1,q^2,v^1,v^2,q^2,q^1) \textrm{ such that } q^A,\, v^A\in\mathbb{R}\}=\textrm{graph}(FL)$.

Next, we compute 
\[
\begin{array}{ll}
W_2=&\{(q^1,q^2,v^1,v^2,q^2,q^1)\in W_1 \textrm{ such that there exists }\\ \noalign{\medskip}& X\in T_{(q^1,q^2,v^1,v^2,q^2,q^1)}W_1  \textrm{ satisfying } i_{X}\, \Omega=dD\}
\end{array}
\]

If $X\in TW_1$ then $X$ can be locally expressed as
\begin{equation}\label{condition2}
\begin{array}{l}
X=a^1 \frac{\partial}{\partial q^1}+a^2 \frac{\partial}{\partial q^2}+b^1 \frac{\partial}{\partial v^1}+b^2 \frac{\partial}{\partial v^2}\\ \noalign{\medskip}
+ (\frac{\partial^2 L }{\partial v^1\partial q^1 }a^1+\frac{\partial^2 L }{\partial v^1\partial q^2 }a^2+\frac{\partial^2 L }{\partial v^1\partial v^1 }+\frac{\partial^2 L }{\partial v^2\partial v^1 })\frac{\partial}{\partial p_1}
 \\ \noalign{\medskip}
+(\frac{\partial^2 L }{\partial v^2\partial q^1 }a^1+\frac{\partial^2 L }{\partial v^2\partial q^2 }a^2+\frac{\partial^2 L }{\partial v^2\partial v^1 }b^1+\frac{\partial^2 L }{\partial v^2\partial v^2 }b^2)\frac{\partial}{\partial p_2} 
\\ \noalign{\medskip}
=a^1 \frac{\partial}{\partial q^1}+a^2 \frac{\partial}{\partial q^2}+b^1 \frac{\partial}{\partial v^1}+b^2 \frac{\partial}{\partial v^2}+ a^2 \frac{\partial}{\partial p_1}+a^1 \frac{\partial}{\partial p_2}
\end{array}
\end{equation}

Taking into account  \eqref{conditions} and \eqref{condition2}, for every point $(q^1,q^2,v^1,v^2,q^2,q^1) \linebreak \in W_1$ we obtain
\[
X=v^1\frac{\partial}{\partial q^2}+v^2\frac{\partial}{\partial q^2}+b^1\frac{\partial}{\partial v^1}+b^2\frac{\partial}{\partial v^2}+v^2\frac{\partial}{\partial p_1}+v^1\frac{\partial}{\partial p_2}
\]
for arbitrary $b^1, \ b^2$, and so $W_2=W_1$ and therefore the final constraint submanifold is $W_1$; consequently,  $Q_f=Q$.

Now, a solution of the Hamilton-Jacobi problem in the Skinner-Rusk setting is given by $\sigma=(Z,\gamma)$ such that 
\begin{enumerate}
\item Im$(\sigma)\subset W_1$.
\item Im$(\sigma_f)\subset W_f$.
\item $d(pr_2\circ \sigma)=d\gamma=0$, that is, $\gamma$ is closed.
\item $d(D\circ \sigma)_{|Q_f}=0$
\end{enumerate}

It is easy to see that every pair given by a  vector field $Z$ and its image by the Legendre transformation, that is $(Z,\gamma=FL(Z))$ is a solution of the problem. In fact, by construction Im$(\sigma)\subset W_1$ and $D_{|W_1}=0\Rightarrow D\circ \sigma=0$. Following the argument in example \ref{example} we can take $Z(q^1,q^2)=\frac{\partial}{\partial q^1}+\frac{\partial}{\partial q^2}$, and so 
\[
\sigma(q^1,q^2)=\frac{\partial}{\partial q^1}+\frac{\partial}{\partial q^2}+q^2dq^1+q^1dq^2
\]

If we consider the solution 
\[
X(q^1,q^2,v^1,v^2)=v^1\frac{\partial}{\partial q^1}+v^2\frac{\partial}{\partial q^2}+\frac{\partial}{\partial v^1}+\frac{\partial}{\partial v^2}+v^2\frac{\partial}{\partial p_1}+v^1\frac{\partial}{\partial p_2}
\]
then 
\[
X^{\sigma}(q^1,q^2)=\frac{\partial}{\partial q^1}+\frac{\partial}{\partial q^2}
\]
and
\[
T\sigma(X^{\sigma}(q^1,q^2))=\frac{\partial}{\partial q^1}+\frac{\partial}{\partial q^2}+\frac{\partial}{\partial p_1}+\frac{\partial}{\partial p_2}.
\]
A direct inspection shows that
\[
T\sigma(X^{\sigma}(q^1,q^2))\neq X(\sigma(q^1,q^2)=\frac{\partial}{\partial q^1}+\frac{\partial}{\partial q^2}+\frac{\partial}{\partial v^1}+\frac{\partial}{\partial v^2}+\frac{\partial}{\partial p_1}+\frac{\partial}{\partial p_2}.
\]
}
\end{example}

\vspace{0.3cm}
We can also obtain information of the Hamilton-Jacobi problem in the Skinner-Rusk setting from a solution of the Hamilton-Jacobi problem in the hamiltonian side.

If $\gamma$ is a solution of the Hamilton-Jacobi problem in the hamiltonian setting and $Y$ a vector field on $M_f$ wich is a solution of equation \eqref{hamilton}, then we can define $Y^{\gamma}$ as before.

We can also define a section $\tilde{\sigma}$ of $pr_f:W_f\rightarrow
Q_f$ given by $\tilde{\sigma}(p)=(Y^{\gamma}(p),\gamma(p))$ for all $p\in Q_f$. An easy
computation shows that $T\tilde{\sigma}(Y^{\gamma})$ is a vector field
along Im($\tilde{\sigma}$) which solves
\eqref{equation}. Moreover if we find a vector field $Z$ on $Q$ such
that $FL\circ Z=\gamma$ and $Z_f=Y^{\gamma}$, then the pair
$(Z,\gamma)$ is a solution of the Hamilton-Jacobi problem in the
Skinner-Rusk setting.

\section{Appendix: The Gotay-Nester-Hinds algorithm of constraints}
In this section we will briefly review the constraint algorithm of constraints for presymplectic systems (see \cite{gotay6,gotaythesis}).

Let $M_1$ be a  manifold, $\Omega$ a presymplectic structure on $M_1$, i.e., $\Omega$ is a closed $2$-form, and $\alpha$ a $1$-form on $M_1$. We will call $(M_1,\ \Omega,\ \alpha)$ a presymplectic system. 

Gotay {\it et al.} developed an algorithm to find $N$, a submanifold  of $M_1$ where we can solve the equation
\begin{equation}\label{ecu}
i_X\, \Omega=\alpha
\end{equation}
with $X$ tangent to $N$.

The previous equation could not hold for every point of $M_1$, because $\alpha$ could not be in the range of $\Omega$. So it is necesary to introduce the following set
\[
M_2=\{p\in M_1 \textrm{ such that there exists } X\in T_{p}M_1 \textrm{ satisfying }i_X\, \Omega=\alpha\},
\]
and it is assumed that $M_2$ is a submanifold.

At the points of $M_2$ there exists solution to equation \eqref{ecu} but in an algebraic sense, that is, the solution could not be tangent to $M_2$. This forces a further restriction to
\[
M_3=\{p\in M_2 \textrm{ such that there exists } X\in T_{p}M_2 \textrm{ satisfying }i_X\, \Omega=\alpha\},
\]
which is also assumed to be a submanifold.

Proceeding as above, the algorithm will produce a sequence of submanifolds
\[
\cdots M_3 \ldots \hookrightarrow^{j_3}  M_2 \hookrightarrow^{j_2} M_1
\]
where
\[
M_{l+1}=\{p\in M_l \textrm{ such that there exists } X\in T_{p}M_{l} \textrm{ satisfying }i_X\, \Omega=\alpha\},
\]
and $j_l$ denote the inclusions.

There are three possibilities:
\begin{enumerate}
\item There exists $k$ such that $M_k=\O$.
\item There exists $k$ such that $M_k=M_{k+1}$.
\item The algorithm does not end.
\end{enumerate}

In the second case the submanifold $M_k$ is called the final constraint submanifold and is denoted by $M_f$. By construction there exists a vector field on $M_f$ such that is solution of equation \eqref{ecu}. The third case is only possible in the infinite dimensional setting. In this case, the final constraint submanifold is defined by $M_f=\cap_{i=1}M_i$.

Note that the final constraint submanifold is maximal in the sense that if $R$ is submanifold of $M_1$ where there exists a tangent solution of equation \eqref{ecu}, then $R\subset M_f$.

\section*{Acknowledgments}
This work has been partially supported by MICINN (Spain) \linebreak MTM2010-21186-C02-01, the European project IRSES-project ``Geo\-mech-246981'' and the ICMAT Severo Ochoa project SEV-2011-0087.

\end{document}